# On the Power of Impersonation Attacks

Michael Okun*
Weizmann Institute of Science,
Rehovot 76100, Israel

**Abstract**

In this paper we consider a synchronous message passing system in which in every round an external adversary is able to send each processor up to $k$ messages with falsified sender identities and arbitrary content. It is formally shown that this impersonation model is slightly stronger than the asynchronous message passing model with crash failures. In particular, we prove that $(k+1)$-set agreement can be solved in this model, while $k$-set agreement is impossible, for any $k \geq 1$. The different strength of the asynchronous and impersonation models is exhibited by the order preserving renaming problem, for which an algorithm with $n + k$ target namespace exists in the impersonation model, while an exponentially larger namespace is required in case of asynchrony.

## 1 Introduction

In the standard message passing models, whether synchronous or asynchronous, with crash, omission or Byzantine failures, it is typically assumed that the identity of a message sender is known to the receiver. However, in practice, due to impersonation attacks by malicious adversaries, this often is not the case. Various attacks of this kind have been extensively investigated in the past, in particular for peer-to-peer and sensor networks [4, 8]. However, these studies were done in the context of network security or cryptography, rather than distributed computing theory. In the present work, we study one particular model of impersonation attacks, as part of ongoing research on the theoretical implications of relaxing the assumption on foolproof identification of the message senders [9, 10].

We consider a set of $n$ processors, $p_1, ..., p_n$, communicating by message passing. Instead of requiring some specific communication topology (such as a fully connected network), we only assume that every $p_i$ is able to send a message to any other $p_j$, and that the message sender is identified by including its id in the message. For simplicity the communication is assumed to be synchronous. The adversary is an external entity capable of injecting messages with arbitrary content into the network, but it cannot prevent the processors from receiving each other's messages. The ids of the processors are assumed to be fixed and known a priori, thus injecting messages that impersonate the real processors is the only way by which the adversary can interfere with the computation. For example, if two messages tagged by $p_1$ are received, it might be impossible to know which message is the real one. Adversarial behavior of this kind is known as *stolen identities Sybil attack* [4, 8]. For the purpose of formal analysis, the strength of the adversary is quantified by the number of messages it is able to send to each processor in every round. A *k-adversary* can generate up to $k$ additional messages for every processor, so that a processor can receive up to $n + k$ messages in a round, instead of the usual $n$ (including the message sent to oneself).

---

*Email: michael.okun@mail.huji.ac.il

At first we have considered the consensus problem. There exists a simple FLP-type proof [5] for the impossibility of deterministic consensus, even in the presence of 1-adversary. On the other hand the problem can be solved with probability 1 in presence of $k$-adversary for $n > 2k$, using the paradigm proposed by Ben-Or in [2]. In addition to possible implications for systems where impersonation attacks are of concern, these results are interesting from theoretical viewpoint, because of the close analogy to the asynchronous crash-failure model.

The analogy, suggested by the above results, is not accidental. Recall that an asynchronous computation resilient to $t$ crashes, when presented in the *normal form* (e.g., see [6]), consists of rounds, where in each round the adversary removes $t$ of the messages sent to each processor. Thus, it is possible to simulate an asynchronous environment with at most $t$ crash failures on top of a $t$-adversarial impersonation model in the following manner. Whenever two (or more) messages tagged with the same source id $p_i$ arrive, they are dropped. Since at most $t$ messages can have fake "twins", messages from at most $t$ processors are dropped, exactly as in the asynchronous computation.

The opposite direction, however, does not work - the $k$-adversary model is strictly stronger than the asynchronous model with $k$ failures, because all the messages sent in every round by the processors are received. For example, in the impersonation model each processor is able to compute (in a single round) an upper bound on the input values of all the processors, which is impossible in the asynchronous case with even a single failure.

To further investigate the relationship between the impersonation and asynchronous models, we considered the other two central distributed coordination tasks - renaming and $k$-set agreement.

For the renaming problem we present a simple order-preserving renaming algorithm resilient against a $k$-adversary, that has an optimal target namespace of size $n + k$. In the asynchronous case the minimum possible size of the target namespace of any order preserving algorithm resilient to $t$ failures is $2^t(n - t + 1) - 1$ [1]. This further exemplifies the difference between the two models. The result also shows that various impossibility results shown for the asynchronous model are due to different reasons. For asynchronous order preserving consensus the large target namespace is a result of complete uncertainty about the input values of some processors. In the impersonation model this uncertainty is reduced - eventually the input of each processor is known to belong to a small set of possible values, and as a result the size of the target namespace is significantly smaller. For consensus it is important to know the exact input value of each processor, thus it is impossible both in asynchronous model with single crash failure and in the impersonation model with 1-adversary.

In the $k$-set agreement problem the values of at least $n + k - 1$ processors have to be known exactly, therefore its behavior is similar in both asynchronous and impersonation models. In more detail, there exists an algorithm in presence of $(k-1)$-adversary, but no deterministic algorithm resilient against a $k$-adversary. Since the impersonation model is stronger than the asynchronous model, this fact cannot be directly inferred from the known results. Nevertheless, the proof methods originally developed for the asynchronous model can be applied to the impersonation model, just as in the case of consensus. Therefore, to derive the $k$-set agreement lower bound we use the combinatorial topology machinery developed in [6, 3].

When put together, our results show that the effects of impersonation attack on a synchronous message passing system and the loss of synchrony are very much alike. The subtle difference in the computational power of the two models is not evident for consensus and $k$-set agreement. On the other hand, renaming, which is the easiest among the three coordination problems, reveals that the models are not equivalent.

The rest of the paper is structured as follows. Section 2 deals with the consensus problem. In Section 3 we present an order preserving algorithm for $n > k^2 + k$ with target namespace of size $n + t$, resilient to $k$-adversary. It is also shown that this result is optimal. Section 4 presents a $k$-set agreement algorithm resilient to $(k − 1)$-adversary, and a proof that no such algorithm exists for $k$-adversary.

## 2 Consensus

In this section we show that consensus cannot be solved even in the presence of 1-adversary. We provide two different proofs of this fact. The first one is an FLP-style bivalency argument. The second proof involves an explicit construction of a similarity chain between executions where all the inputs are 0 and 1, respectively. This construction is later used in Section 4, where the $k$-set agreement problem is considered. In addition, we show that the problem can be solved with probability 1 in the presence of $k$-adversary for $n > 2k$.

### 2.1 Bivalency proof

DEFINITION. A $r$-round execution $\mathcal{E}$ of an algorithm $\mathcal{A}$ is called *univalent* if all the possible continuations of $\mathcal{E}$ have the same decision value [7]. A univalent execution is called 1-valent if the only possible decision value is 1, and is called 0-valent if the only possible decision value is 0. If $\mathcal{E}$ is not univalent, it is called *bivalent*.

**Lemma 2.1** *There exists an initial bivalent configuration.*

**Proof.** An initial configuration is described by a vector in $\{0, 1\}^n$, which specifies the input of every processor $p_i$ ($1 \leq i \leq n$). If every such configuration is univalent, then there exist configurations $C_0$ and $C_1$ which differ in exactly one coordinate $1 \leq i_0 \leq n$, such that $C_0$ is 0-valent and $C_1$ is 1-valent. Let $\mathcal{E}_0$ be an execution whose initial configuration is $C_0$, in which the adversary acts throughout the execution as $p_{i_0}$ with input as given in $C_1$. Similarly, let $\mathcal{E}_1$ be an execution whose initial configuration is $C_1$, in which the adversary acts $p_{i_0}$ with input specified from $C_0$. The two executions are indistinguishable to any processor other than $p_{i_0}$. Therefore, the decision is the same in both. However, by the assumptions, the decision value in $\mathcal{E}_0$ is 0 and the decision value in $\mathcal{E}_1$ is 1, which is a contradiction. □

Next, using the same idea we show that any bivalent $r$-round execution has a bivalent extension in round $r + 1$.

**Lemma 2.2** *Let $\mathcal{E}$ be a bivalent $r$-round execution of the algorithm $\mathcal{A}$, where $r \geq 0$. Then there exists a $(r + 1)$-round bivalent execution $\mathcal{E}'$ which is an extension of $\mathcal{E}$.*

**Proof.** The proof is presented in Appendix A.

From Lemma 2.1 and Lemma 2.2 it follows that any algorithm $\mathcal{A}$ has infinite executions. Thus we have proved the following theorem.

**Theorem 1** *In the presence of 1-adversary there exists no deterministic algorithm that solves the consensus problem.*

## 2.2 Similarity chain proof

Below we provide an explicit construction of a similarity chain between $R$-round failure-free (ff) executions (i.e., executions where the adversary sends no messages) in which all the inputs are 0 and 1, respectively (for any $R > 0$). This shows that there exists no consensus algorithm.[1]

In each round of every execution in the chain, the adversary will impersonate at most one correct processor . Moreover, it will do so only by mimicking an admissible state of that processor. Formally, for any given execution this state can be characterized as follows.

DEFINITION. The *inverse* of the state of a processor $p_i$ in the beginning of round $r$ is defined as follows. For $r = 1$, the input value of $p_i$ is the opposite of its input in the original state. For $r > 1$, if the original state of $p_i$ reflects receiving $n + 1$ messages in round $r - 1$, where one of them was sent by the adversary, then the state of $p_i$ had it not received this message is defined as the inverse. Similarly, if originally $p_i$ did not receive a message from the adversary in round $r - 1$, then the inverse state reflects the case in which $p_i$ does receive it. However, in case the state of the adversary is not defined at all in round $r - 1$, the inverse is equal to the original. ∎

Using this definition we can describe executions by means of *communication graphs*, somewhat similarly to the way it is done in the standard crash failure model. A communication graph consists of $(n+1) \times (R+1)$ grid of vertices ($\{<p_i, r>\}_{1 \leq i \leq n,\ 0 \leq r \leq R}$ and $\{<Adv, r>\}_{0 \leq r \leq R}$), each representing one processor or the 1-adversary in a specific round. Since in our model the processors always exchange messages, it will be convenient not to represent these messages in the graph. Thus, the only edges in the communication graph are between vertices of the form $<Adv, r-1>$ and $<p_i, r>$, where $1 \leq r \leq R$, and $1 \leq i \leq n$. Such an edge represents a message sent by the adversary to $p_i$ in round $r$.

To describe the state of the adversary, the $<Adv, \cdot>$ vertices in the graph are labeled by numbers $1, ..., n$ or by the special $A$ label. The meaning of label $1 \leq i \leq n$ assigned to the vertex $<Adv, r>$ is that in the execution defined by the graph, the state of the adversary in the beginning of round $r + 1$ is the inverse of $p_i$'s state. The $A$ label means that the state of the adversary is derived from its own state in the previous round. More specifically, if it was impersonating $p_i$ in round $r - 1$, then its state in the end of round $r$ is computed according to the algorithm for $p_i$, where the messages received in round $r$ are those received by $p_i$ in that round. If there is no label for $<Adv, r>$, then the state of the adversary in the beginning of round $r + 1$ is undefined, and it must not send any messages in this round.

A communication graph must satisfy the following restrictions. A vertex $<Adv, r>$ cannot be labeled by $A$ unless $<Adv, r-1>$ is also labeled by $A$ or by some $1 \leq i \leq n$. In particular, $< Adv, 0 >$ can be labeled only by $1 \leq i \leq n$. In addition, if $< Adv, r >$ is labeled by $A$, then $< Adv, r-1 >$ must be connected to $< p_1, r-1 >, ..., < p_n, r-1 >$. This is required because the state of the adversary in the beginning of round $r+1$ is computed from its state in the previous round, as if it was some processor, therefore it must behave in round $r$ like a correct processor, i.e., all its messages must be received.

---

[1]At first it may seem unclear why showing that there exists no algorithm with an a priori known bound on the running time also proves the impossibility of consensus. This follows from the observation that for any consensus algorithm there must exist a bound on its running time (as was already noted in [6], page 865). To see this, we can consider the whole protocol as a game (and thus a tree) played by the adversary, where in each step there is a finite number of possible moves. A tree without infinite paths which has only finite degree vertices cannot have an unbounded depth.

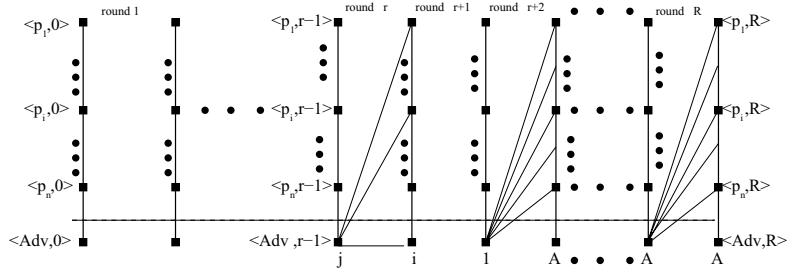

Figure 1: An example of a communication graph.

Defined this way, a communication graph provides a complete description of an execution of the algorithm $\mathcal{A}$. An example graph is depicted in Fig. 1. For the formal proof we use the following (standard) definitions.

DEFINITION. Two $R$-round communication graphs $\mathcal{G}$ and $\mathcal{G}'$ of some algorithm $\mathcal{A}$ are *similar* if $n-1$ (out of the total $n$) processors have the same view in the executions defined by them. The two graphs are *similarity connected* if there is a chain of similar graphs that starts with $\mathcal{G}$ and ends with $\mathcal{G}'$. Let $\sim$ denote the similarity relation and let $\approx$ denote the transitive closure of $\sim$ (similarity connectedness). ∎

**Theorem 2** *For any $R > 0$, there exists a similarity chain between a ff communication graph with all 0 inputs and a ff graph with all 1 inputs.*

**Proof.** The structure of the proof reminds similar proofs for the synchronous crash failure model. The proof is presented in Appendix A.

### 2.3 Randomized algorithm
In Section 1 we described a simple way to simulate any algorithm for asynchronous system with $t$ crash failures on top of a system with impersonations by $t$-adversary. When applied to the algorithm proposed by Ben-Or in [2], we get an algorithm that achieves consensus with probability 1, for $n > 2t$.

## 3 Renaming
In the renaming problem each processor gets a unique natural number as an input, and has to output a unique value from some a priori fixed set $M$. The algorithm should be independent of the processor ids $p_1, ..., p_n$ [1], e.g., by having an identical code for all the processors. In addition, here the processor

```
Initialization:
  get the input value v_o
  V := ∅
In round 1:
  send v_0 to all
In round 2:
  FOR any value v received from p
    send echo'(p,v) to all
  IF echo'(p,v) was received from n distinct processors
    send in the next round echo(p,v) to all
In round 3,...,n+k+4
  FOR every (p,v) ∈ V
    send echo(p,v) to all
  IF in the present round echo(p,v) was received from n distinct sources
    V := V ∪ {(p,v)}
  IF in the present round echo(p,v) was received from n-k distinct sources
    send in the next round echo(p,v) to all
```

Figure 2: A protocol for constructing the vector $V$.

ids are considered as entities which can be tested for equality but not compared (i.e., $p = q$ is either true or false, while $p > q$ is undefined). Thus renaming is based solely on the natural number input values.

This section presents an order-preserving renaming algorithm with target namespace of size $n+k$, resilient against $k$-adversary. It works for $n > k^2 + 2k$ and requires $n+k+4$ rounds.[2] The algorithm has two layers. The first layer (Fig. 2) uses the standard echoing technique in order to maintain a vector $V$ of pairs $(p, v)$, where $p$ is a processor id and $v$ is the associated input value. Due to the presence of the adversary, $V$ may contain more than one pair with the same processor id.

The next lemma establishes the properties of this vector that are required for the renaming algorithm.

**Lemma 3.1** *For any processor $p_i$ with input $v_i$, the pair $(p_i, v_i)$ belongs to the vector $V$ of every processor by the end of the third round. Also, if $(p, v)$ belongs to the vector $V$ of $p_i$ in round $r \geq 3$, then by the end of round $r+1$, $(p, v)$ belongs to the vectors of all the processors. Finally, for every processor it holds that $|V| \leq n+k$.*

**Proof.** The proof is rather standard and appears in Appendix B. □

The renaming algorithm, presented in Fig. 3, is layered on top of the protocol which computes the vector $V$. Note that it does not send any messages of its own.

**Theorem 3** *The algorithm in Fig. 3 solves the order-preserving renaming problem. Furthermore, the size of its target namespace, $n+k$, is optimal*

**Proof.** First we argue that every processor decides. A simple induction on $r \geq 1$ shows that if a processor did not decide in round $r + 3$, then the rank of its input value among all the values in its

---
[2] For smaller values of $n$ the $k$-adversary is able to enact more than $k$ processors. This aspect of the problem, which has no analogs in the asynchronous crash failure model, is not considered here.

```
In the end of round r + 3 (n + k ≥ r ≥ 1):
  S := {v|∃ p s.t. (p,v)) ∈ V}
  IF not decided yet AND r = rank(v_0, S) both in the end of the present
  and the previous rounds, THEN
    DECIDE r
```

Figure 3: Renaming algorithm.

$V$-vector is higher than $r$. Since according to the previous lemma the size of $V$ is bounded by $n + k$, the processor has to decide by round $(n + k) + 3$.

Now, suppose by contradiction that two processors $p_i$ and $p_j$ decide on the same new name $r$. W.l.o.g., assume that $v_i < v_j$, where $v_i$ and $v_j$ are the input values of $p_i$ and $p_j$, respectively. According to Lemma 3.1, every value in the set $S$ of $p_i$ in round $r + 2$ must belong to the set $S$ of $p_j$ in round $r + 3$, contradicting the fact that $v_i$ and $v_j$ have the same rank in both sets. The same argument shows that the algorithm is order preserving.

To show that target namespace of size $n+k$ is optimal for any order-preserving renaming algorithm in the presence of $k$-adversary, consider an execution in which there are two instances of $p_1$ with inputs 1 and $n + 1$, two instances of $p_2$ with inputs 2 and $n + 2$, ..., two instances of $p_k$ with inputs $k$ and $n + k$, and processors $p_{k+1},...,p_n$ with inputs $k + 1,...,n$, respectively. Since any of the two copies of $p_1,...,p_k$ can be attributed to the adversary, any $p_i$ and $p_j$ where $i \neq j$ must output different values. Moreover, the decision value of $p_i$ with input $i$ and the decision value of $p_i$ with input $n + i$ (for $1 \leq i \leq k$) are different, because the first one must be lower than the decision of $p_n$ while the second one must be higher. □

## 4 $k$-set agreement

In the $k$-set agreement problem each processor has an input value from a set $V$ and has to decide so that the cardinality of the set of output values of all the processors does not exceed $k$ and so that its decision is equal to an input value of some processor. When the set $V$ is large, every processor might have a distinct input value. In this case it is impossible to distinguish between an input value of some processor and an input value presented by the $k$-adversary which behaves like a correct processor. To circumvent this (artificial) difficulty, we will assume that the number of processors $n$ satisfies $n > |V|k$, in which case it is sufficiently large to ensure that at least one input value can be distinguished from the values introduced by the $k$-adversary (by counting the number of its instances).

### 4.1 An algorithm

There is a simple 2-round algorithm for $(k+1)$-agreement in presence of $k$-adversary, which is presented in Fig. 4.

**Lemma 4.1** *For $n > |V|k$ the algorithm in Fig. 4 solves the $(k + 1)$-agreement problem in the presence of $k$-adversary.*

**Proof.** First, observe that any input value that is common to at least $k+1$ processors must appear in the set $M$ of all the processors. In particular, this implies that the set $M$ of every processor contains at least one value.

```
Initialization:
 Let v_0 be the input value from V
 M := ∅
In round 1:
 send v_0 to all
In round 2:
 FOR every message v received in round 1 from more than k sources
   send echo(v) to all
 RECEIVE MESSAGES
 FOR every message v received in round 1 from more than k sources
   IF echo(v) was received from n distinct sources
     M := M ∪ {v}
 DECIDE min(M)
```

Figure 4: $k$-set agreement algorithm.

Let $v$ be a value that belongs to the set $M$ of some processor $p_i$, but not to the set of $p_j$. It follows that $n - k$ processors sent an $echo(v)$ message in the second round. Each of these processors must have received a message with $v$ from the adversary in the first round. Since the total number of messages that the processors can receive from the adversary in the first round is bounded by $nk$, the number of values that appear in the set $M$ of some correct processors, but are absent in others, is bounded by $nk/(n-k) < k+1$. The inequality follows from $n > |V|k \geq (k+1)k$. Therefore the set of all the decision values cannot exceed $k+1$ (otherwise two of the decision values appear in the sets of all processors, which is impossible given the way they are chosen). □

## 4.2 Impossibility result

In this section we show that there is no $k$-set agreement algorithm for the $k$-adversary case for $V = \{0, ..., k\}$. Assume by contradiction that there exists an algorithm $\mathcal{A}$ which solves the problem in $R$ rounds (as noted in Section 2 such an a priori bound must exist). Let $\sigma_i$ denote the similarity chain between a communication graph where all the inputs are 0 and a graph where all the inputs are $i$ $(1 \leq i \leq k)$, for the case of 1-adversary, (constructed in Section 2). Let $N+1$ be the length of these chains, and let $\mathcal{G}_i(j)$ denote $j$th graph in $\sigma_i$ $(0 \leq j \leq N)$.

The structure of the proof is similar to the lower bound proof for $k$-set agreement in the synchronous crash failure model [3]. The topological part of the proof is significantly simplified by the $n \gg k$ assumption, which is later shown to be unessential (by a simple simulation argument).

Let $B$ denote a $k$-dimensional simplicial complex (the Bermuda triangle) which is the Kuhn's triangulation of the simplex defined by $v_0 = (0, 0, ...0)$, $v_1 = (N, 0, ...0)$, $v_2 = (N, N, ...0)$,..., $v_k = (N, N, ..., N)$, (see [7, 3] for more details). First, each vertex $x = (x_1, ..., x_k)$ of $B$ is labeled by a communication graph $\mathcal{G}(x)$ which describes an execution in the presence of $k$-adversary. This graph is formed by merging the communication graphs $\mathcal{G}_1(x_1), ..., \mathcal{G}_k(x_k)$. We will show that if a vertex $x$ belongs to the convex hull of $v_{j_1}, ..., v_{j_l}$, then all the processors in $\mathcal{G}(x)$ decide on values that belong to $\{j_1, ..., j_l\}$. In the next stage, every $x \in B$ is labeled by a processor $p(x) \in \{p_1, ..., p_n\}$. By assuming $n \gg k$ we can ensure that the view of $p(x)$ is the same in all the executions in the neighborhood of $x$. Next, we assign each $x \in B$ a color between 0 and $k$, which is equal to the decision of $p(x)$ in $\mathcal{G}(x)$. Because of the above mentioned property of executions assigned to vertices on the faces of $B$, this is a Sperner coloring. By Sperner's lemma there must exist a simplex in $B$, whose vertices, $x^0, ..., x^k$, have distinct colors. This implies that $p(x^0), ..., p(x^k)$ decide on $k+1$ distinct values in $\mathcal{G}(x^0)$ (as well as in $\mathcal{G}(x^1), ..., \mathcal{G}(x^k)$), which is a contradiction. Finally, we argue that showing the impossibility for $n \gg k$ is sufficient, since an algorithm for $n_0$ would have implied an algorithm for any $n \geq n_0$.

For the formal proof, we define communication graphs for describing executions in the presence of $k$-adversary. These graphs have the same general structure as in the 1-adversary case (in Section 2), with several modifications described below. A graph describing an execution in presence of $k$-adversary contains $k$ nodes that represent it ($<Adv_1, r>, ..., <Adv_k, r>$, where $0 \leq r \leq R$). As before, every such node represents an admissible state of some processor, enacted by the adversary. Label $i$ on node $<Adv_j, r>$ has the following meaning. If $r = 0$, then the input value for the instance of $p_i$ simulated by the adversary is defined to be $j$ when the input value of the real $p_i$ is different from $j$, and $j - 1$ when the input value of the real $p_i$ is equal to $j$. If $r > 0$ and $<Adv_j, r-1>$ is connected to $<p_i, r>$, then the state of $<Adv_j, r>$ is defined to be the state the real $p_i$ would have had, had it not received the message represented by the edge between $<Adv_j, r-1>$ and $<p_i, r>$. The case in which $<Adv_j, r-1>$ is not connected to $<p_i, r>$ is defined correspondingly. The $A$ label has the same meaning as in the 1-adversary case.

The communication graph $\mathcal{G}(x)$ assigned to vertex $x = (x_1, ..., x_k)$ is formed by merging the communication graphs $\mathcal{G}_1(x_1), ..., \mathcal{G}_k(x_k)$. The merge operation is defined in the intuitive way - the label and connections of $<Adv_i, r>$ are defined according to the $<Adv, r>$ node in the $i$-th graph. The input value for $p_i$ ($1 \leq i \leq n$) is defined to be $p_i$'s maximal input value over all the graphs. This assignment has several important properties, shown in the following lemmas.

**Lemma 4.2** *If $x$ belongs to a $(l-1)$-dimensional face of $B$ which is the convex hull of $v_{j_1}, ..., v_{j_l}$, then all the correct processors in $\mathcal{G}(x)$ decide on values that belong to $\{j_1, ..., j_l\}$.*

**Proof.** W.l.o.g. suppose $j_1 < j_2 < ... < j_l$. Let $0 \leq i \leq k$ be some number such that $i \notin \{j_1, ..., j_l\}$. To prove the claim it is sufficient to show that no correct processor in $\mathcal{G}(x)$ has input $i$. Suppose $j_{a-1} < i < j_a$ for some $1 \leq a \leq l$. Since $x$ belongs to the convex hull of $v_{j_1}, ..., v_{j_l}$, $x_i = x_{j_a}$. The symmetry between chains implies that every correct processor which has input $i$ in $\mathcal{G}_i(x_i)$, must have input $j_a$ in $\mathcal{G}_{j_a}(x_{j_a})$. By definition of $\mathcal{G}(x)$ the claim follows.

If $i < j_1$, then $x_i = x_{j_1}$ and the claim follows by the same argument. If $i > j_l$, then the $i$-th coordinate of $x$ is equal to 0, so that no processor in $\mathcal{G}_i(x_i)$ has $i$ as input value in the first place. □

**Lemma 4.3** *Let $x = (x_1, ..., x_k)$ be some vertex of $B$, such that $x' = (x_1, ..., x_{i-1}, x_i+1, x_{i+1}, ..., x_k) \in B$ (where $1 \leq i \leq k$). Then $\mathcal{G}(x) \sim \mathcal{G}(x')$.*

**Proof.** The proof is presented in Appendix C (Note that it depends on the technicalities of the similarity chain construction, as described in Section 2 and Appendix A). □

Next we assign each vertex $x$ of $B$ a correct processor $p(x) \in \{p_1, ..., p_n\}$. Such an assignment induces a $(k+1)$-coloring of $B$, by taking the decision value of $p(x)$ in $\mathcal{G}(x)$ to be the color of $x$.

DEFINITION. Two vertices $x$ and $x'$ of $B$ are *neighbors* if $|x_i - x'_i| \leq 1$ for every $1 \leq i \leq k$.

**Lemma 4.4** *For sufficiently large $n$ there exists an assignment of processors to vertices of $B$ such that for any two neighbors $x, x' \in B$ the view of $p(x)$ in $\mathcal{G}(x)$ and in $\mathcal{G}(x')$ is the same.*

**Proof.** The proof is presented in Appendix C. □

**Theorem 4** *There exists no $k$-set agreement algorithm in presence of $k$-adversary, for $n > k(k+1)$.*

**Proof.** We assign each vertex $x$ of $B$ a color which is equal to the decision value of $p(x)$ in the execution described by $\mathcal{G}(x)$. According to Lemma 4.2 this is a Sperner coloring, and thus Sperner's lemma implies that there exists a simplex colored by all $k+1$ colors. Thus the processors assigned to the vertices of this simplex decide on $k+1$ different values in each one of the executions corresponding to the vertices of the simplex. However, by Lemma 4.4 the view of each among these processor is the same across the simplex, which contradicts the initial assumption on the algorithm $\mathcal{A}$.

Finally we argue that the $n \gg k$ assumption made in Lemma 4.4 is not necessary. Suppose that there exists a $k$-set agreement algorithm for $n_0 > k(k+1)$. It can be transformed into an algorithm for arbitrarily large $n$ in the following manner. First the processors $p_1,...,p_{n_0}$ execute the original algorithm. After completing it, each processor sends its decision to all the processors. Processors $p_{n_0+1},...,p_n$ decide on any value that appears in more than $k$ messages. □

# Appendix A

**Proof of Lemma 2.2.** An extension of $\mathcal{E}$ to the next round is defined by a $n$-dimensional vector $M = (m_1, ..., m_n)$ of messages, where $m_i$ is the message that the adversary sends to $p_i$. If all the extensions of $\mathcal{E}$ are univalent, then there exist two message vectors $M_0$ and $M_1$ which differ in exactly one coordinate, such that the $(r+1)$-round execution defined by $M_0$ is 0-valent, while the one defined by $M_1$ is 1-valent. Assume w.l.o.g. that $M_0$ and $M_1$ differ in the message sent to $p_1$. Let $\mathcal{E}'_0$ denote the $(r+1)$-round extension of $\mathcal{E}$ defined by $M_0$, and let $\mathcal{E}'_1$ denote the $(r+1)$-round extension of $\mathcal{E}$ defined by $M_1$.

Now, define $\mathcal{E}_0$ as an extension of $\mathcal{E}'_0$ in which the adversary acts as $p_1$ which received in round $r+1$ the message specified by $M_1$, starting from round $r+2$. Similarly, define $\mathcal{E}_1$ as an extension of $\mathcal{E}'_1$ in which the adversary acts as $p_1$ which received in round $r+1$ the message specified by $M_0$, starting from round $r+2$. The executions $\mathcal{E}_0$ and $\mathcal{E}_1$ are indistinguishable to any correct processor other than $p_1$, yet the decision value in $\mathcal{E}_0$ is 0 and the decision value in $\mathcal{E}_1$ is 1, which is a contradiction. $\square$

In the rest of this appendix we present the technical part of the proof of Theorem 2. We start with two definitions.

DEFINITION. A graph $\mathcal{G}$ is $r$-*failure free* ($r$-*ff*) if there are no labels on $<Adv, r>, ..., <Adv, R>$. ∎

DEFINITION. A graph $\mathcal{G}$ is a $<p_i, r>$-*graph* if $<Adv, r>$ is labeled by $i$ and $<Adv, r+1>, ..., <Adv, R>$ are labeled by $A$. ∎

Every graph in the similarity chain is formed from its predecessor by one of the three transformations defined below. The first two transformations change the state of the processor enacted by the adversary, in a way that is invisible to the correct processors. The last transformation switches between a processor and its "alter ego" created by the adversary, starting from some round $r$ till the end.

$label(\alpha, r)$  An operation which labels $<Adv, r>$ in any $r$-ff graph with $\alpha \in \{A, 1, ..., n\}$.

$remove(r)$  Removes a label from $<Adv, r>$, provided $<Adv, r>$ is not connected to any vertex of the form $<p_i, r+1>$. This operation is provided mainly for convenience, since we could have left this label until the next $label(\cdot, r)$ operation.

$switch(r)$  This operation can be applied only if the graph is a $<p_i, r>$-graph for some $p_i$. In this case, in rounds $r+1,...,R$ all the processors cannot tell which of the two messages having $p_i$ as their source id comes from the correct processor. The operation switches between the states of the processor enacted by the adversary and $p_i$ in rounds $r+1, ..., R$. The operation is invisible to all the other correct processors. Whenever $<Adv, r-1>$ is labeled, the operation adds or removes the edge between $<Adv, r-1>$ and $<p_i, r>$. For $r=0$ this operation changes the input value of $p_i$.

**Lemma.** *Let $\mathcal{G}_{ff}$ be some $r$-ff graph. Let $\mathcal{G}_{p_i}$ denote the $<p_i, r>$-graph formed from $\mathcal{G}_{ff}$ by labeling $<Adv, r>$ with $i$, $<Adv, r+1>, ..., <Adv, R>$ with $A$ and connecting all these vertices to all the* *correct processors.* Then $\mathcal{G}_{ff} \approx \mathcal{G}_{p_i}$, *moreover the chain between these two graphs can be constructed using the three above defined operations.*

**Proof.** The proof is by reverse induction on $r$. For $r = R$ the claim is trivial. Next, we prove the claim for $r < R$, using the fact that it is true for $r + 1$. The similarity connectedness of $\mathcal{G}_{ff}$ and $\mathcal{G}_{p_i}$ is shown in several steps:

(i) Let $\mathcal{G}'$ denote the graph accepted from $\mathcal{G}_{ff}$ by the $label(i, r)$ operation. Obviously, $\mathcal{G}' \sim \mathcal{G}_{ff}$.

(ii) According to the induction assumption $\mathcal{G}' \approx \mathcal{G}'_{p_1}$, where $\mathcal{G}'_{p_1}$ denotes a graph accepted from $\mathcal{G}'$ by labeling $<Adv, r+1>$ with 1, $<Adv, r+2>$, ..., $<Adv, R>$ with $A$ and connecting all these vertices to all the correct processors. The graph $\mathcal{G}'_{p_1}$ is presented in Fig. 1. Now we apply the $switch(r+1)$ operation which connects $<Adv, r>$ with $<p_1, r+1>$. Finally, since the graph remains to be a $<p_1, r+1>$-communication graph, we apply the induction assumption once again, to remove the labels from $<Adv, r+1>, ..., <Adv, R>$.

(iii) The procedure described in (ii) is repeated for $p_2, ..., p_n$, to get an execution where $<Adv, r>$ is connected to $<p_1, r+1>, ..., <p_n, r+1>$. Now $<Adv, r+1>$ can be labeled with $A$.

(iv) Steps (ii)-(iii) are repeated in rounds $r+2$,...,$R$ in order to label the $<Adv, r+2>$,..., $<Adv, R>$ with $A$. □

**Proof of Theorem 2.** Let $\mathcal{G}$ be some 0-ff graph, i.e., a specification of input values. By the previous lemma, using the same notation, $\mathcal{G} \approx \mathcal{G}_{p_i}$ for any $p_i$. Since all the correct processors except for $p_i$ cannot distinguish between the two copies of $p_i$, we can switch between the correct $p_i$ and the one enacted by the adversary, and then apply the lemma again to remove the faulty copy of $p_i$. This argument shows that any ff graph where $p_i$'s input is 0, is similarity connected to a ff graph where its input is 1. Since the same is true for every processor, the theorem follows. □

# Appendix B

**Proof of Lemma 3.1.** Suppose that $(p, v)$ belongs to the set $V$ of processor $p_i$ in round $r$. It follows that at least $n$ $echo(p, v)$ messages were received, and at most $k$ of them were sent by the adversary. From the algorithm it directly follows that every processor sends an $echo(p, v)$ message in round $r+1$, and thus $(p, v)$ is added to $V$ of every processor in round $r+1$ at the latest.

Next we prove the $n + k$ bound on the size of $V$. Observe that $(p, v)$ belongs to the set $V$ of some processor iff at least $n - k$ processors send an $echo'(p, v)$ message in round 2. It follows that $v$ was received in the first round by at least this number of processors. The total number of messages sent by the adversary to the processors in the first round is bounded by $nk$. Therefore the number of values $v$ in $V$ that are not the actual input values is bounded by $nk/(n - k) < k + 1$. The last inequality follows from the $n > k^2 + k$ assumption. □

# Appendix C

**Proof of Lemma 4.3.** The differences between $\mathcal{G}(x)$ and $\mathcal{G}(x')$ result from the differences between $\mathcal{G}_i(x_i)$ and $\mathcal{G}_i(x_i+1)$, which in turn are a result of applying one of the *label*, *remove*, *switch* operations (defined in Appendix A) to the former graph. It is easy to verify that for each of these operations the claim holds. The only nontrivial case is when an input value of some processor $p_l$ changes between $\mathcal{G}_i(x_i)$ and $\mathcal{G}_i(x_i + 1)$, since different graphs interact with each other in the merge operation by overwriting lower input values with higher ones. For this case observe that $p_l$ has the same input value in both $\mathcal{G}(x')$ and $\mathcal{G}_i(x_i + 1)$, because $x_i + 1 > x_j$ for any $j > i$. This implies the similarity of $\mathcal{G}(x)$ and $\mathcal{G}(x')$. $\square$

**Proof of Lemma 4.4.** Define $y = (\max(0, x_1 - 1), \max(0, x_2 - 1), ..., \max(0, x_k - 1))$. Let $S = \{z \in B | y_i \leq z_i \leq y_i + 2 \ \forall \ 1 \leq i \leq k \}$. Clearly, $|S| \leq 3^k$ and all the neighbors of $x$ belong to $S$. From Lemma 4.3 it follows that there are at most $3^k$ processors whose view is not the same in all the graphs assigned to vertices in $S$. Thus, whenever $n > 3^k$ a processor as required must exist. $\square$